\documentclass[twocolumn]{aastex62}
\usepackage{amsmath}
\usepackage{graphicx}
\usepackage{natbib}
\usepackage[scaled]{helvet}
\usepackage{epsfig}
\usepackage{url}
\usepackage{textcomp}
\usepackage{mathpazo}
\usepackage{xspace}
\usepackage{hyperref}
\usepackage{apjfonts}
\usepackage{mathrsfs}
\usepackage{verbatim}
\usepackage{color,colortbl}
\usepackage[normalem]{ulem}
\usepackage{multirow}
\usepackage{lineno}

\definecolor{Gray}{gray}{0.9}

\bibpunct{(}{)}{;}{a}{}{,}
\newcommand{\feh}{\ensuremath{\left[{\rm Fe}/{\rm H}\right]}}
\newcommand{\initfeh}{\ensuremath{\left[{\rm Fe}/{\rm H}\right]_0}}

\newcommand{\teff}{\ensuremath{T_{\rm eff}}}

\newcommand{\esinw}{\ensuremath{e\sin{\omega_{*}}}}
\newcommand{\lcosw}{\ensuremath{L\cos{\omega_{*}}}}
\newcommand{\lsinw}{\ensuremath{L\sin{\omega_{*}}}}
\newcommand{\vcve}{\ensuremath{V_c/V_e}}
\newcommand{\secosw}{\ensuremath{\sqrt{e}\cos{\omega_{*}}}}
\newcommand{\sesinw}{\ensuremath{\sqrt{e}\sin{\omega_{*}}}}

\newcommand{\mplanet}{\ensuremath{\,M_{\rm P}}}
\newcommand{\rplanet}{\ensuremath{\,R_{\rm P}}}

\newcommand{\cosi}{\ensuremath{\cos{i}}}

\newcommand{\kepler}{Kepler}
\newcommand{\ktwo}{K2}
\newcommand{\tess}{TESS}
\newcommand{\gaia}{Gaia}

\newcommand{\mstar}{\ensuremath{M_{*}}}
\newcommand{\rstar}{\ensuremath{R_{*}}}
\newcommand{\ar}{\ensuremath{a/R_*}}

\newcommand{\exofasttwo}{{\tt EXOFASTv2}}

\newcommand{\tonefour}{\ensuremath{T_{14}}}
\newcommand{\tfwhm}{\ensuremath{T_{\rm FWHM}}}

\newcommand{\tc}{\ensuremath{T_{\rm C}}}

\newcommand{\tmps}{\ensuremath{T_{\rm T}}}
\newcommand{\tp}{\ensuremath{T_{\rm P}}}

\begin{document}

\title{A Novel Eccentricity Parameterization for Transit-only Models}

\author[0000-0003-3773-5142]{Jason D.\ Eastman}
\affiliation{Center for Astrophysics \textbar \ Harvard \& Smithsonian, 60 Garden St, Cambridge, MA 02138, USA}

\shorttitle{Eccentricity re-parameterization}
\shortauthors{Eastman}

\begin{abstract}
We present a novel eccentricity parameterization for transit-only fits that allows us to efficiently sample the eccentricity and argument of periastron, while being able to generate a self-consistent model of a planet in a Keplerian orbit around its host star. With simulated fits of 330 randomly generated systems, we demonstrate that typical parameterizations often lead to inaccurate and overly precise determinations of the planetary eccentricity. However, our proposed parameterization allows us to accurately -- and often precisely -- recover the eccentricity for the simulated planetary systems with only transit data available.

\end{abstract}



\section{Introduction}

Despite the unquestioned success of \kepler, \ktwo, and \tess, the abundance of candidates and faintness of their hosts is a significant problem for follow up and characterization with the limited high precision radial velocity facilities, leaving the majority of known exoplanets without measured masses or eccentricities. 

The planetary eccentricity's impact on the transit lightcurve was first described by \citet{Tingley:2005}. \citet{Barnes:2007} noted that we could determine a lower limit on the eccentricity from the transit photometry using an independent constraint on the star, which was expanded on by \citet{Ford:2008}, and later applied to a real system as the ``photoeccentric effect'' \citep{Dawson:2012}. It is also the same basic idea behind ``astrodensity profiling'' \citep{Kipping:2012} and used to argue about the ensemble eccentricities of certain populations of planets \citep{VanEylen:2015}. All of these methods model the transiting planet assuming a circular orbit, derive a constraint on the stellar density from the lightcurve \citep{Seager:2003}, and compare it to the known stellar density to determine the planetary eccentricity.

With the advent of several all-sky surveys that provide quality, absolute, broad-band photometry across the stellar spectrum and precise, astrometric parallaxes from \gaia, we can use Spectral Energy Distribution (SED) and evolutionary modeling to determine independent stellar densities for nearly all exoplanet host stars -- often only limited by systematic errors \citep{Tayar:2022} -- making this technique widely applicable.

However, assuming the planet's orbit is circular creates an inconsistency between the properties of the star and the properties of the planet that orbits it, biasing the the value of \ar. This is an important, fundamental property that affects the semimajor axis, the equilibrium temperature, transit probabilities, and occurrence rates. This method is embodied in the ensemble analyses of the complete set of Kepler Objects of Interest (KOIs) and K2 Candidates \citep{Mayo:2018,Thompson:2018} because the true constraint on the eccentricity from the lightcurve is degenerate and difficult to sample.

Indeed, we will show that using naive parameterizations of the eccentricity that are common when including radial velocities leads to inaccurate and overly precise eccentricities from the lightcurve, despite passing convergence criteria commonly used in the literature to determine the quality of Markov Chain Monte Carlo fits.

Here we present a more efficient parameterization of the planetary eccentricity that allows us to create a self-consistent global model of the star and planet, efficiently constrain the eccentricity only using the transit light curve, and generate an accurate eccentricity posterior that is often surprisingly precise. We also validate the accuracy and precision of our models by fitting 330 simulated transit light curves.

\section{Degeneracy between $\MakeLowercase{e}$ and $\omega_*$}

The eccentricity $e$ and argument of periastron $\omega_*$\footnote{See \citet{Eastman:2019} for the complete definition of this widely confused parameter} is most commonly parameterized as \secosw \ and \sesinw, which reduces the covariance between $e$ and $\omega_*$, eliminates the problematic angular parameter $\omega_*$, and naturally imposes a uniform prior on both $e$ and $\omega_*$ \citep{Anderson:2011,Eastman:2013}. When we fit radial velocities or astrometry, the eccentricity and argument of periastron are well constrained and this parameterization is well behaved\footnote{It is worth emphasizing that this parameterization does not remove the Lucy-Sweeney bias \citep{Lucy:1971}, which observes that the model eccentricity can only scatter upward from the hard boundary at zero, biasing $e$ to nominally higher values. As a result, in order to be 95\% confident the eccentricity is non-zero, it must be $2.45\sigma$ above zero -- not the naive value of $2.0\sigma$.}.

\begin{figure}[!htbp]
  \begin{center}
    \includegraphics[width=3.5in, trim={0.5cm 0cm 1.5cm 0cm}]{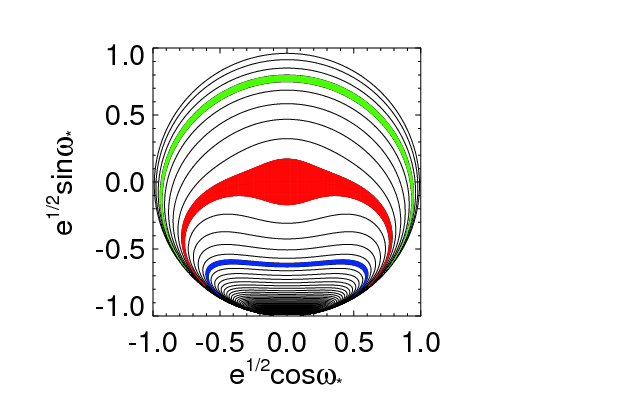} 
    \caption{The covariance between \secosw \ and \sesinw \ for transit-only fits, shown with contours in equal transit duration, only including solutions with $0 \leq e < 1$. The shaded regions denote the typical constraint from the transit duration consistent with a circular orbit (red), half that duration (green), or 1.5 times that duration (blue). We see that the vast majority of parameter space is eliminated, including preferentially eliminating high eccentricity solutions. We can also see the diabolically nonlinear covariance that is difficult to sample. Differential evolution or Affine invariant samplers will draw a significant number of steps in the unlikely regions inside the contours, severely impacting their efficiency.}
    \label{fig:secoswsesinw}
  \end{center}
\end{figure}

Unfortunately, the \secosw \ and \sesinw \ parameterization has a diabolical covariance when fitting a transit and stellar density alone, owing to the fact that essentially, we have one constraint (the transit duration) to constrain two physical parameters ($e$ and $\omega_*$) and the translation between the physical parameters and the constraint is not straight-forward. Figure \ref{fig:secoswsesinw} shows the covariance between \secosw \ and \sesinw \ with contours of constant transit duration (assuming a fixed inclination, stellar density, and orbital period). The shaded regions denote the typical constraint from the transit duration consistent with a circular orbit (red), half that duration (green), or 1.5 times that duration (blue). This covariance is inefficient for Differential Evolution (DE) or Affine Invariant (AI) Markov Chain Monte Carlo (MCMC) algorithms to sample. These algorithms essentially draw two random points from the Probability Distribution Function (PDF) that define a vector to draw the next step. For linear covariances, that vector is a natural means of efficiently sampling that covariance. For curved covariances, like in figure \ref{fig:secoswsesinw}, it often leads to proposed steps in the low-likelihood regions interior to the curve, causing the acceptance rate -- and therefore the efficiency of the MCMC -- to plummet ($\lesssim 1\%$ is typical for transit-only fits parameterized as \secosw \ and \sesinw -- $\sim20\times$ less than ideal). More sinister, this covariance is particularly difficult to fully explore, and can often pass less strict convergence criteria without properly sampling the tips of the covariance between \secosw \ and \sesinw, biasing the eccentricity toward its starting value and underestimating the uncertainties in both $e$ and $\omega_*$.

\section{Reparameterizing $\MakeLowercase{e}$ and $\omega_*$}
\label{sec:edegeneracy}

To make sampling this diabolical degeneracy more efficient, we can reparameterize $e$ and $\omega_*$. Ideally, we would diagonalize the fisher matrix of the entire global model to make all parameters independent of one another. However, \citet{Carter:2008} showed that, even with a significant number of simplifying approximations, the ideal parameterization of just the transit depends on the parameters themselves. That is, no such general parameterization exists.

Fortunately, the DE-MCMC and AI-MCMC algorithms work well in the regime of linear covariances, so the parameters need not be uncorrelated, just linearly correlated on the scale of a typical constraint.

\begin{figure}
  \begin{center}
    \includegraphics[width=3.5in]{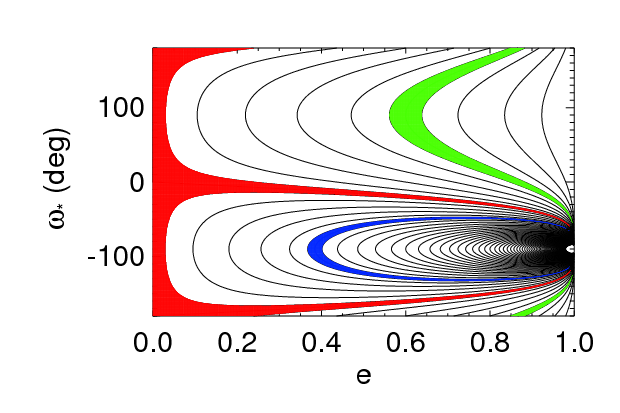} 
    \caption{The covariance between $e$ and $\omega_*$, as shown with contours in equal $V_c/V_e$ for transit-only fits. The shaded regions are the same as Figure \ref{fig:secoswsesinw}.}
    \label{fig:eomega}
  \end{center}
\end{figure}

Unfortunately, the most straight-forward choice, parameterizing in $e$ and $\omega_*$ directly, is perhaps even more diabolically degenerate, as shown in Figure \ref{fig:eomega}.

So let us first consider the observed quantity, the transit duration from first to fourth contact, $T_{14}$, which is approximately equal to:

\begin{equation}
\label{eq:t14}
    T_{14} \approx \frac{P}{\pi}\arcsin{\left[\frac{R_*}{a}\frac{C}{\sin{i}}\right]}\frac{\sqrt{1-e^2}}{1+\esinw},
\end{equation}

\noindent where $C$ is the transit chord, equal to

\begin{equation}
    \label{eq:chord}
    C =\sqrt{(1+R_P/R_*)^2-b^2},
\end{equation}

\noindent and $b$ is the impact parameter equal to 

\begin{equation}
    \label{eq:b}
    b =\frac{a}{R_*}\cos{i}\frac{1-e^2}{1+\esinw}.
\end{equation}

Unfortunately, the arcsin in equation \ref{eq:t14} requires an expensive and complex numerical approach to both deriving the physical parameters necessary to generate a model and computing the Jacobian to correct the implicit priors to be physical.

Instead, the transit duration can be parameterized as a unitless scaling factor between the velocity the planet would have if its orbit were circular, $V_c$, divided by the velocity of the planet at the time of transit in its eccentric orbit, $V_e$, which \citet{Winn:2010} showed is approximately equal to

\begin{equation}
\label{eq:vcve}
    \frac{V_c}{V_e} \approx \frac{\sqrt{1-e^2}}{1+\esinw}.
\end{equation}

Comparing equation \ref{eq:t14} with equation \ref{eq:vcve}, we see that we have entirely ignored the duration's dependence on the period, inclination, and planet size. This makes it far more manageable, and dramatically simplifies the covariance in well-constrained fits, but means that in regimes where the period, inclination, or planet size is not well-constrained (such as single or grazing transits), this parameter is a poor substitute for the observed transit duration and remains highly degenerate with the observed quantities.

While this quantity is only approximately equal to the velocity, it is important to recognize that this approximation does not impact the accuracy of our model. It is merely a tool to step through parameter space, from which we derive the precise values of the parameters and generate the physical model without approximating the planet's velocity during transit.

However, there are several subtle problems this reparameterization introduces, which we address in the following subsections.

\subsection{Transit chord}

The constraint from the observed transit duration only scales nicely with the planet's velocity, \vcve, when the transit chord, $C$, is known. Unfortunately, the transit chord, even with a fixed $a/R_*$ and $i$, changes as a function of $e$ and $\omega_*$ (and therefore \vcve), meaning our standard \cosi \ parameterization leaves a covariance between \cosi \ and \vcve \ similar to the one we were trying to avoid, and limits the improvement in performance. Therefore, it is also helpful to reparameterize the orbital inclination as the transit chord instead of \cosi. Using $b$ instead of \cosi \ or $C$ would be a more conventional parameterization and would also remove the dependence on $e$ in the translation from the parameterization to the observed constraint. We did not attempt this because we expect $C$, the distance traveled, to be more closely related to the observed duration and therefore better behaved, especially when coupled with the planet's velocity.

For large swaths of parameter space, $V_c/V_e$ and $C$ are linearly correlated on the scale of a typical constraint. Even in the worst cases, the covariance is no more diabolical than the original parameterization. The result is that we accept more proposed steps, explore parameter space more thoroughly, and converge faster.

\subsection{Deriving $\omega_*$}
\label{sec:omega}

In order to derive the physical parameters $e$ and $\omega_*$ from $V_c/V_e$, we must introduce another parameter. Fitting for $\omega_*$ directly is an obvious choice, but angular parameters are periodic that create several difficulties with MCMC algorithms. Formally, periodic parameters can never converge, because the likelihood is identical at integer multiples of the period. Even in practice, it is sometimes possible for multiple chains to get stuck in separate minima, making it practically impossible to converge. Rejecting steps near an artificial boundary could bias the posterior.

Choosing just \esinw \ or just $\sin{\omega_*}$ would allow us to solve for $e$, but not $\omega_*$ in its full $2\pi$ range. This can be generally solved by reparameterizing a single angle $\omega_*$, into two fitted parameters, \lcosw \ and \lsinw, bounded such that $\left(\lcosw\right)^2 + \left(\lsinw\right)^2 = L \leq 1$. L must be bounded to allow convergence, and the bound must be circular to recover a uniform prior in $\omega_*$, though the choice of 1 is arbitrary. Then, we compute $\omega_* = {\rm atan2}\left(\lsinw,\lcosw\right)$. Note that fitting two parameters with one constraint necessarily introduces a degeneracy, but it is perfectly linear (everywhere along the line from the center to the edge of the unit circle has equal likelihood), which is well-handled by modern MCMC algorithms.

Therefore, we sample in both \lsinw \ and \lcosw, and marginalize over the new, meaningless parameter $L$.

\subsection{Sign of the quadratic solution}
\label{sec:sign}

Further, when we solve Equation \ref{eq:vcve} for $e$, it is a quadratic,

\begin{equation}
    \label{eq:equadratic}
      \begin{split}
          0 =& \left[\left(\frac{V_c}{V_e}\right)^2\sin{\omega_*}^2 + 1\right]e^2 + \\        &\left[2\left(\frac{V_c}{V_e}\right)^2\sin{\omega_*}\right]e + \\        &\left[\left(\frac{V_c}{V_e}\right)^2 - 1\right],
\end{split}
\end{equation}

\noindent meaning there are two solutions for $e$ given values of $V_c/V_e$ and $\omega_*$. We must somehow choose between the solution that uses the positive sign and the solution that uses the negative sign. 

\begin{figure}[!htbp]
  \begin{center}
    \includegraphics[width=3.5in]{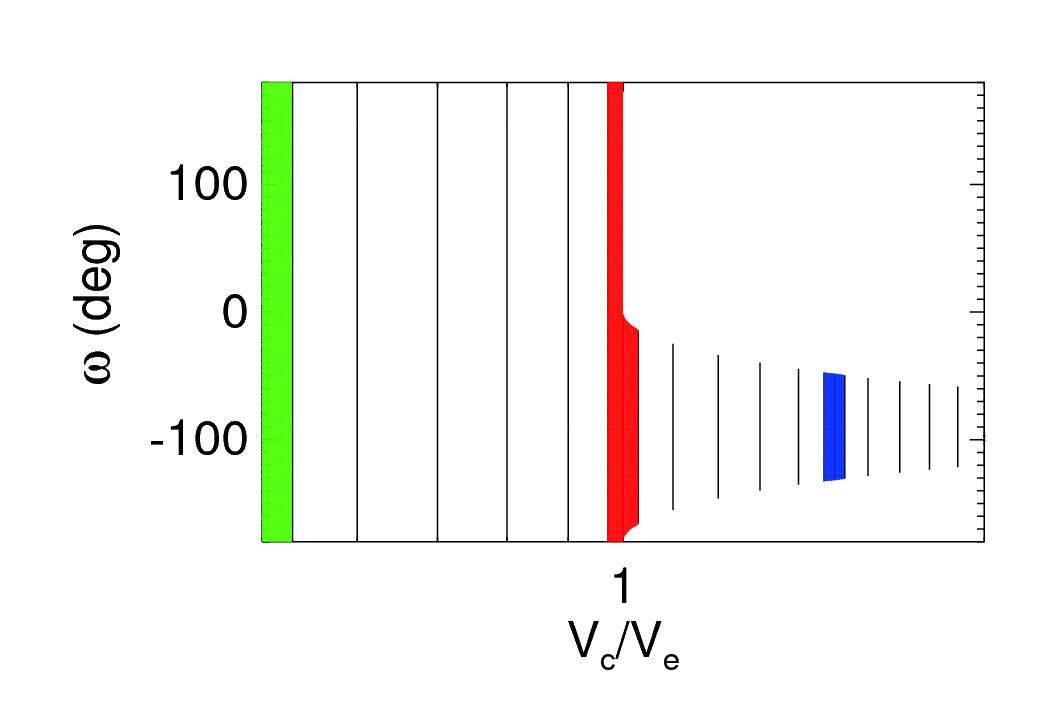} 
    \caption{The covariance between $V_c/V_e$ and $\omega_*$, as shown with contours in equal $V_c/V_e$ for transit-only fits. The shaded regions are the same as Figure \ref{fig:secoswsesinw}. The X-axis is log spaced.}
    \label{fig:vcve}
  \end{center}
\end{figure}

Because $L$, when using \lsinw, \lcosw, is a totally degenerate quantity, we tried using $L$ to choose between the two quadratic solutions and avoid introducing another parameter, but the convergence times were slower. See Appendix \ref{sec:elparam} for a more detailed discussion, but for this reason, we advocate fitting an additional sign parameter, $S$, to choose between these solutions. We bound $0 \leq S < 2$, and when $S<1$, we choose the solution with the positive sign.

\subsection{\tmps \ vs \tc}
\label{sec:tt}

With high eccentricities and inclinations, the time of conjunction \tc, often used as the transit time and defined as the time the true anomaly is $\pi/2 - \omega_*$, can differ by more than 10 minutes from the observed quantity \tmps, the time of minimum projected separation between the planet and star, as seen by an observer on Earth \citep{Eastman:2019}. We normally prefer stepping in \tc \ because the time of periastron \tp, required to compute the model, is trivially computed from \tc. However, for these eccentric, inclined orbits, the covariance between \tc, \vcve, \ and $C$ is curved and inefficient to sample. It is well worth the computational overhead to numerically compute \tp \ from the fitted parameter \tmps \ at each step \ to avoid this complex covariance.

\subsection{\mplanet \ parameterization}
\label{sec:mp}

The planet mass, \mplanet \ is most often parameterized as the observed RV semi-amplitude, $K$, which we cannot measure for the transit-only fits we describe here. In cases where the RV data is available, it should be included in the model, as the additional expense of computing the RV model is small relative to the dramatic decrease in runtime required to properly sample the e-omega degeneracy, even with our re-parameterization proposed here.

Still, we advocate explicitly modeling \mplanet \ anyway, so we can create a self-consistent physical model without having to complicate our more general global model with edge cases that determine which combination of data sets constrain what parameters necessitating various approximations. Instead, we always fit for some parameter related to \mplanet \ and require the user to explicitly place priors to constrain it, e.g., from the \citet{Chen:2017} mass--radius relation or impose an explicit assumption that the planet mass is negligible/zero.

In order to determine the inclination from the transit chord, we must know $a/R_*$. Since we derive it from Kepler's law, \mstar, \rstar, and the planet period rather than fit it, we must know or neglect the planetary mass. But computing the mass from the normally fitted RV semi-amplitude requires the inclination, setting up a nasty system of equations to solve. Instead of neglecting the planet mass or solving that system of equations, we reparameterize the RV semi-amplitude as the planetary mass, \mplanet, which also allows us to impose a more physical prior. For simplicity and to impose intuitive priors, we advocate always fitting in \mplanet, regardless of the parameterization or data being fit.

\subsection{Priors}

The final complication is that, as with any non-physical parameterization, we must be careful about the implicit prior it imposes. The uniform step in $V_c/V_e$ imposes a non-physical prior that strongly biases $e$ toward high eccentricities. Similarly, the uniform step in the transit chord imposes a non-physical prior that strongly biases the orbital inclination toward grazing transits. 

We must correct for these priors by weighting the likelihood of the step by the absolute value of the Jacobian of the transformation between the two parameterizations. In general, this Jacobian is the absolute value of the determinant of a square matrix where the $i$th column and $j$th row is equal to $\partial X_i/\partial Y_j$, $X$ is an array of the parameterized variables, and $Y$ is an array of the parameters we wish to have uniform priors. When most of the parameters are identical or uncorrelated between $X$ and $Y$, the determinant dramatically simplifies, and in our case, is much simpler than we might have feared:

\begin{equation}
    \label{eq:eccjacobian}
    \left|\frac{\partial V_c/V_e}{\partial e} \frac{\partial C}{\partial \cos{i}}\right| = \left|\frac{e+\sin{\omega_*}}{\sqrt{1-e^2}(1+e\sin{\omega_*})^2} \frac{b^2}{cos{i}{C}}\right|.
\end{equation}

To confirm our corrected implicit priors reproduce our expected physical priors, we must first figure out what we expect. By using the transit chord as a parameter, we implicitly exclude non-transiting systems (for which the transit chord is imaginary) in a way that cannot be corrected.\footnote{Similarly, had we used $\sqrt{1-b^2}$ as the transit chord, as is sometimes advocated for its simplicity, we would a priori exclude $b>1$ grazing transits, which also can cannot be corrected.} Fortunately, imposing such a prior is desired because it is a real selection effect that must be accounted for when characterizing the system. That is, eccentric systems are more likely to transit \citep{Burke:2008}, and so, all else being equal, a planet we detect via transit should a priori be expected to be more eccentric than a planet detected via RVs. Such a prior is also practically required because it avoids the infinite volume of parameter space with identical likelihood (a non-transiting model light curve is a flat line for all planets) that is impossible for an MCMC to reasonably explore. In fact, whenever we fit a transit light curve, we explicitly add this limit anyway to avoid sampling the infinite volume of non-transiting parameter space.

However, whether or not the planet transits depends on $a/R_*$, (which itself is derived from the planetary period, \mstar, \rstar, and \mplanet), \cosi, $e$, and $\omega_*$, and so imposing a prior that the planet must transit necessarily skews all of those priors away from uniform, even if they were sampled directly. That is the behavior we want, but makes it difficult to verify our corrected priors are sensible because they are not trivially uniform.

So first, we must figure out what priors we actually expect. We created a dummy model where we fit the parameters $e$, $\omega_*$, and $\cos{i}$, which are the parameters we expect to be biased by our reparameterization. We also fit $M_P$, $\log{M_*/M_\Sun}$, $R_*$, $\log{P}$, and $R_P/R_*$ so we can derive the impact parameter. In this simplified fit, the likelihood function is constant except when it exceeds our imposed boundaries for each parameter of $0 < e < 1$, $-\pi < \omega_* < \pi$, $0 < \cos{i} < 1$, $0.5 < M_*/M_\Sun < 2$,  $1 < M_P/M_\Earth < 300$, $0.5 < R_* < 2$, $0.5 < P/days < 100$, and $0.01 < R_P/R_* < 0.1$, where the likelihood is set to 0 (and the step is always rejected). We then run our MCMC over this function. Thus, the posteriors we generate with no further constraints produce our priors. 

\begin{figure*}
\centering
\includegraphics[height=1.55in, trim={0.5cm 0cm 0cm 0.5cm}]{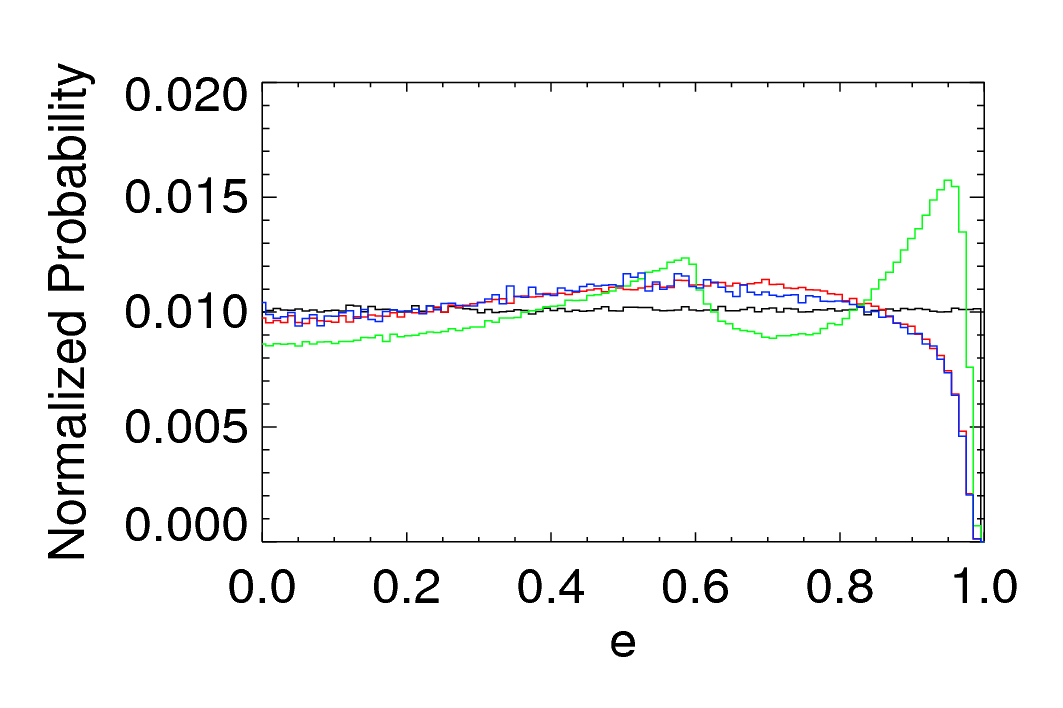}\label{fig:eprior}
\includegraphics[height=1.55in, trim={0.5cm 0cm 0cm 0.5cm}]{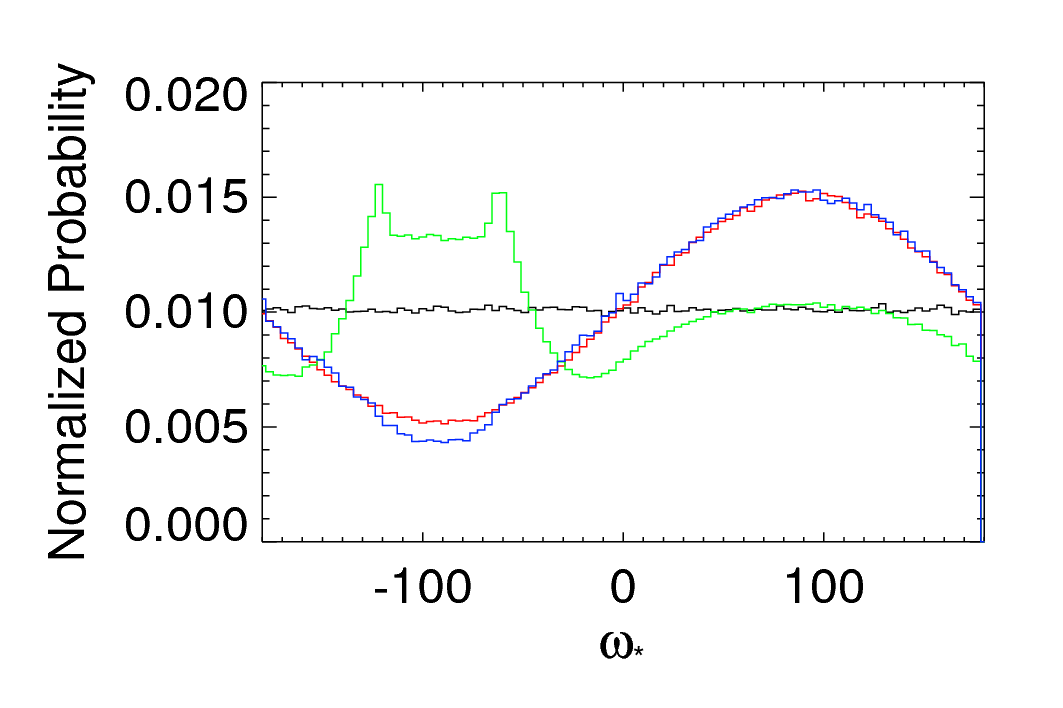}\label{fig:wprior}
\includegraphics[height=1.55in, trim={0.5cm 0cm 0cm 0.5cm}]{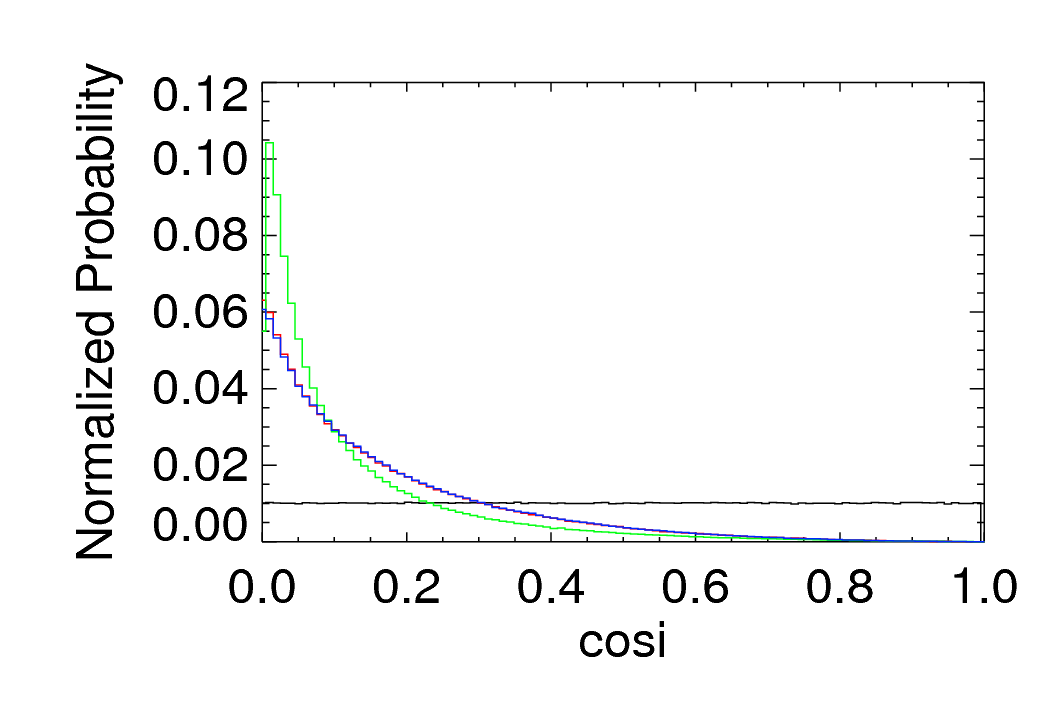}\label{fig:cosiprior}
\caption{(a) -- The implicit prior $e$ for an unconstrained fit that is uniform in $\cos{i}$, \secosw, and \sesinw \ (black), a fit that is uniform in $\cos{i}$, \secosw, and \sesinw, but rejecting non-transiting systems (red), a fit that is uniform in transit chord, $V_c/V_e$, \lsinw, and \lcosw (green), and a fit that is uniform in transit chord, $V_c/V_e$, \lsinw, and \lcosw, but corrected by the Jacobian to impose uniform priors (blue). That is, the red line is the desired prior, and its agreement with the blue demonstrates that we are correctly calculating and applying the Jacobian. See the text for details. (b) -- Same as (a), but for $\omega_*$. (c) -- Same as (a), but for $\cos{i}$. }
\label{fig:ewipriors}
\end{figure*}

The parameters we are most concerned about are $e$, $\omega_*$, $\cos{i}$, and in this simple example, they are stepped in directly and uniformly bounded, and so are trivially uniform, as shown as a black line in Figure \ref{fig:ewipriors}. This fit is just a sanity check to show we are doing things correctly and it behaves as we expect.

Next, we create an identical fit except we also reject steps that do not transit, $b = a/R_*\cos{i}(1-e^2)/(1+e\sin{\omega_*}) < 1+R_P/R_*$. We also have all the necessary information to impose the limit that the star and planet should not collide during periastron, so we also reject steps where $e > 1 - (R_P+R_*)/a$. This now includes the detection bias from $e$, $\omega_*$, and $\cos{i}$ and is shown as a red line in Figure \ref{fig:ewipriors}. These are the desired priors we wish to preserve. We note that the particular distribution of the priors shown here depends on the detailed bounds of the other parameters described above and is only intended as a demonstration. While these bounds were chosen to roughly span the planets \tess \ is sensitive to, it should not be considered a general result, and these bounds should be far less restrictive in a fit with actual data.

Then, we do the same fit, but stepping in our proposed re-parameterization: \vcve, \lcosw, \lsinw, $C$, $S$, $M_P$, $\log{M_*/M_\Sun}$, $R_*$, $\log{\left(P/\rm {days}\right)}$, and $R_P/R_*$. We impose the same bounds as above after deriving the physical parameters, in addition to a few additional constraints that are required to bound the fit in the physical realm: $\lcosw^2 + \lsinw^2 < 1$, $0 < S < 2$, and $0 < C < 1+Rp/R_*$, $0 < V_c/V_e < 2$. Then we rerun the fit. As expected, the resultant prior, shown in green, does not recover our desired prior (red) due to the change in parameterization.

Finally, we run the same fit, but weight our likelihood by the Jacobian (eq \ref{eq:eccjacobian}). We see that this corrected prior (blue) is the same as our desired prior (red) -- that is, our Jacobian correctly recovers our physical priors. 

\subsection{Reparameterization summary}

In summary, for transit-only fits, we reparameterize \tc, \secosw, \sesinw, \mplanet, and \cosi, as \tmps, \vcve, $C$, \lsinw, \lcosw, $S$, and \mplanet, enabling us to recover the physical parameters easily from well-behaved fitted parameters. Adding two additional, degenerate parameters to the fit may seem like a bad idea, but as long as we can map the fitted parameters to a likelihood, it is not fundamentally problematic, and the DE-MCMC or AI-MCMC algorithms deal with poorly constrained parameters like $L$ and $S$ much better than the curved degeneracy between \secosw \ and \sesinw, resulting in a dramatic improvement to the mixing times and accuracy of the result. 


\begin{table*}[ht]
    \centering
    \begin{tabular}{p{0.6in} p{3.0in} p{1.1in} p{1.25in} p{0.25in}}    \hline\hline    
    
    New Param & Description & Prior & Old Param(s) & Reference\\
    \hline
    $C$\dotfill      & Transit chord & $\mathcal{U}[0,1+\rplanet/\rstar]$ & \cosi & Eq \ref{eq:chord}\\
    $\vcve$\dotfill  & Velocity of a planet in circular orbit around the same star with the same period divided by the velocity the eccentric planet at the time of transit. & $\mathcal{U}\left[0,\sqrt{\frac{2a}{\rstar+\rplanet}-1}\right]$ & \secosw, \sesinw & Eq \ref{eq:vcve}\\
    $\lsinw$\dotfill & A degenerate parameter $L$ times the sin of $\omega_*$ & $\mathcal{U}[(\lcosw)^2-1$, & \secosw, \sesinw & \S \ref{sec:omega}\\
                     &                                                        & $1-(\lcosw)^2]$\\

    $\lcosw$\dotfill & A degenerate parameter $L$ times the cos of $\omega_*$ & $\mathcal{U}[(\lsinw)^2-1$, & \secosw, \sesinw & \S \ref{sec:omega}\\ 
                     &                                                        & $1-(\lsinw)^2]$\\
    $S$\dotfill      & Sign parameter to choose the solution of the quadratic in Equation \ref{eq:equadratic}. & $\mathcal{U}$[0, 2] & \secosw, \sesinw & \S \ref{sec:sign}\\
    $T_T$\dotfill    & The minimum projected separation & $\mathcal{U}$[$\tc-P/2$,$\tc+P/2$] & \tc & \S \ref{sec:tt}\\
    $M_P$\dotfill    & The planet mass & None & $K$ & \S \ref{sec:mp}  \\
\hline
    \end{tabular}
    \label{tab:pars}
    \caption{A summary of the new parameterization compared to the old parameterization for transit-only fits. $\mathcal{U}$[c, d] is a uniform distribution bounded inclusively between $c$ and $d$. Despite being a degenerate quantity, a prior on $0 \leq L \leq 1$ is required to ensure a uniform prior in $\omega_*$ and implicitly applies the priors on \lsinw \ and \lcosw. The upper bound on \vcve \ is set implicitly by Equation \ref{eq:vcve} when $\sin{\omega_*}=-1$ through the limit on $e$ that prevents the planet from colliding with the star, $e \leq 1-\frac{\rstar+\rplanet}{a}$.}
\end{table*}

There may be better ways to reparameterize the eccentricity in transit-only cases, but we note that common parameterizations in terms of transit duration ($T_{FWHM}$ and $\tau$, $T_{14}$ and $T_{23}$) are poor choices owing to uncorrectable priors that a priori exclude physically allowed regions of parameter space and may significantly bias the inferred parameters from grazing transits \citep{Carter:2008} that require significant additional complexity to correct \citep{Gilbert:2022}. Our reparameterization is fundamentally and importantly different than other transit duration parameterizations \citep[e.g.,][]{Tingley:2005, Bakos:2007, Kipping:2010}, because we compute an explicit $e$ and $\omega_*$ at each step and generate a projected Keplerian orbit without approximation, allowing us to recover a realistic constraint on the eccentricity from a transit alone, while also modeling a physical Keplerian orbit around a real star.


\section{Tests with simulated systems}
\label{sec:esim}

To test our ability to recover an accurate and precise eccentricity and demonstrate the advantage of our reparameterization, we simulated 330 planetary systems. Each system had 1 planet, and we randomly drew parameters uniformly distributed between $0 \leq e \leq 1$, $0 \leq \omega_* < 2\pi$, $0.5 \leq M_*/M_\Sun \leq 2$, $-0.5 \leq \initfeh \leq 0.5$, $202 < EEP < 454$, $0.001 \leq M_P/M_J \leq 13$, $\log{\left(3\right)} \leq \log{P} \leq \log{\left(365\right)}$ (systems with a single transit were allowed), and a value for $\cos{i}$ that transits (including grazing transits and accounting for eccentricity). Notably, these simulated systems will over-represent high eccentricity planets, which are not actually distributed uniformly in $e$, and long period planets, which are not actually detected uniformly in $\log{P}$. 
We used the MIST relations to define self consistent simulated values for $R_*$, \teff, and \feh \ based on the randomly drawn parameters, then created a TESS-like light curve, sampled at 2 minute cadence for 1 year with 20 ppm precision (i.e., a bright target in \tess's continuous viewing zone). No correlated noise or data gaps were inserted. The times were converted from the implicit target frame to the ``observed'' (Solar System Barycenter) frame, accounting for the light travel time throughout the planet's orbit. 

Next, we set up fits using \exofasttwo \ \citep{Eastman:2019} as closely as possible to what we would do for blind TESS follow up fit. We excluded the data outside of $\tmps \pm \tonefour$ to speed up the fit. We imposed a prior on $M_P$ and disabled the exoplanet mass--radius relation to avoid a potentially problematic degeneracy with the \citet{Chen:2017} relation for $\sim1 R_J$ planets, but this has a negligible impact on the inferred eccentricity. 

Understanding that a BLS search of the \tess \ data will return reasonably accurate values for the transit time (\tmps), duration (\tfwhm), period ($P$), and depth ($\delta$) of a transit, we use the exact values of those derived from the simulated parameters to initialize the simulated fits. We started each fit at the simulated transit time and $R_P/R_*=\sqrt{\delta}$, which will be roughly correct for non-grazing systems, but systematically small for grazing systems, as one would likely do when modeling a real, unknown system. 

Rather than model the star with a simulated SED or MIST models, we imposed Gaussian priors on \mstar \ and \rstar \ equal to the simulated values with uncertainties of 3\%, simulating a typical (systematics dominated) stellar constraint from spectroscopy and an SED. We also fixed the added variance to 0 and out of transit flux to 1. We started most fits at the simulated period, with $i=90^\circ$, and the value of $V_c/V_e$ that reproduces the observed transit duration.

There are two classes of exceptions to this procedure. First, in the cases where a grazing geometry would require a non-physical eccentricity to reproduce the observed transit duration, we set the starting guess for $e$ to 95\% of its maximum physically allowed value ($e=0.95(1-(R_*+R_P)/a)$). If the transit duration is shorter than the nominal circular orbit, we start $\omega_*$ at $\pi/2$ to minimize the transit duration given $e$, and if the transit duration is longer than the nominal circular orbit, we start $\omega_*$ at $-\pi/2$ to maximize the transit duration given $e$. Then, we assume $b=1$ and derive the starting values for $C$ and $V_c/V_e$. 

Second, for single transit systems, we start at a circular orbit ($e=0$), a central crossing transit ($i=90^\circ$), and start the period at a value to match the transit duration, \tfwhm, 

\begin{equation}
P=\frac{\pi \tfwhm^3G(\mstar+\mplanet)}{4\rstar}. 
\end{equation}

Rather than remove the out of transit baseline as in fits with multiple transits, we include the entire lightcurve so that the out of transit baseline effectively sets a lower limit on the period. The exception to this exception is when the period implied by a circular orbit is already excluded by the out of transit data. Then, we set the period to the minimum allowed by the data and scale \vcve \ as above, including the possibility of modifying the impact parameter for a duration that is still too small. 


We disabled the constraint from the \citet{Claret:2017} limb darkening tables, and fixed $F_0=1$ and the transit variance to 0. We enable parallel tempering, run an unthinned 10,000 step preliminary MCMC, then restart the fit at the best-fit found among all MCMC links to optimize the starting position. We ran each of the final fits for 2.5 days on a super computer with 8 threads each, resulting in fits that typically do not pass our strict convergence criteria, but what we would usually consider reliable. 

We ran each fit two times, only changing the parameterization of the fit -- once with our previous standard parameterization of \secosw, \sesinw, and \cosi \ and again with our new default parameterization for transit-only fits, \vcve, \lcosw, \lsinw, $S$, $C$.\footnote{We actually ran 3 other combinations of parameterizations comparing the tradeoff between \cosi \ vs $C$, and $S$ vs letting $L$ define the sign, but they were clearly inferior.} 

Both parameterizations have statistically significant outliers in eccentricity and had some systems that were particularly poorly mixed (less than 50 independent draws or a Gelman-Rubin Statistic of greater than 1.5). However, the \secosw and \sesinw \ parameterization was far worse -- 55 of 330 ($17\%$) systems had $>3\sigma$ outliers, with the worst at 55 sigma discrepant. These fits typically did not travel far from their starting values and the uncertainties were significantly smaller relative to the fit of the same data with the new parameterization. While 121 ($37\%$) of fits were particularly poorly mixed, the remaining were especially worrisome because they achieved reasonable values for the convergence statistics and showed no obvious signs of bias. This is a cautionary tale for why we must never trust convergence statistics alone. 

In contrast, the new parameterization only had 7 ($2\%$) systems with $>3\sigma$ outliers, the worst outlier of 7.4 sigma, and 7 systems that were poorly mixed. The two biggest outliers were both poorly mixed, single-transit systems, which we might expect to be problematic. However, the remaining 5 failures are all eccentric fits that just so happen to have $\vcve \sim 1$ where we systematically underestimate the eccentricity. We investigated each of these and found that the simulated values happen to lie in narrowly allowed regions of $e-\omega_*$ space. Our sampling did find these solutions, and correctly reported their likelihood given the priors and the $e-\omega_*$ degeneracy. We note that the prevalence of such systems in our simulations is exaggerated by the uniform draw in eccentricity we used to generate the simulated systems. In the real world, we expect low eccentricity systems to be far more common, and so outliers like this to be even rarer occurrences, with an accurate probability reflected in the posteriors.

The acceptance rate for both parameterizations is still relatively poor, but the average 1.3\% acceptance rate with the new parameterization is more than double the average acceptance rate of the old parameterization (0.6\%). In addition, many of the rejected steps with the new parameterization are rejected immediately due to a non-physical eccentricity, without having to compute an expensive model, meaning the new parameterization evaluates $\sim$50\% more models in the same amount of time. As a result, they typically reached much higher levels of mixing in the same amount of time (or mixed to the same level much faster). 

\begin{figure*}
\centering
\includegraphics[height=3.0in, trim={6.0cm 0cm 10.5cm 0.5cm}]{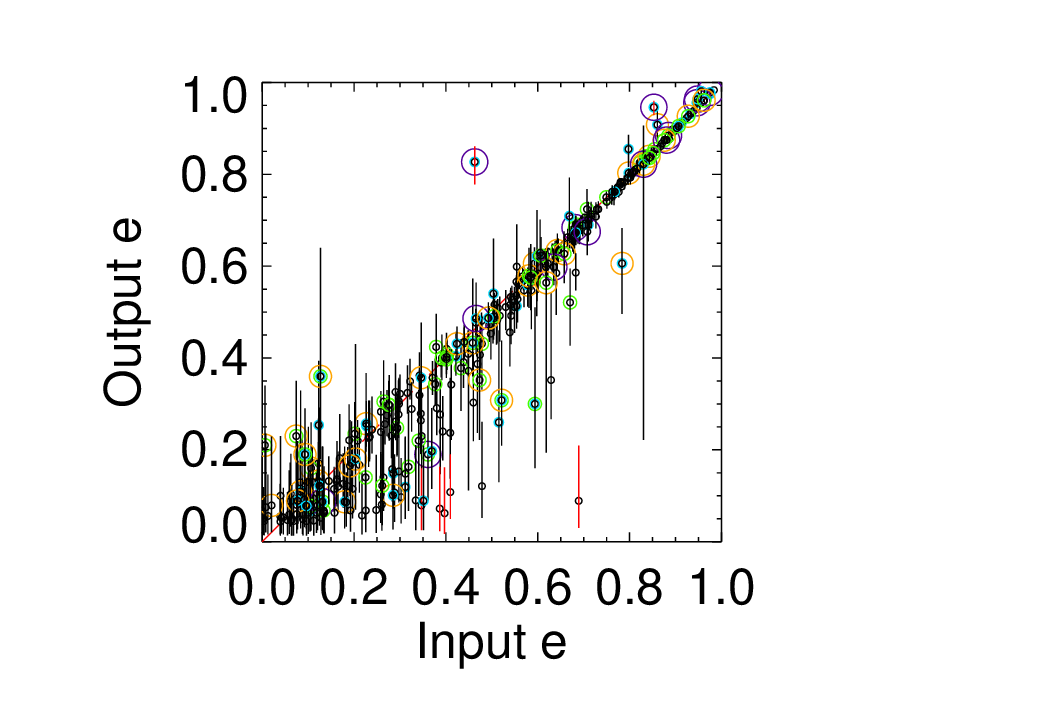}\label{fig:einveout}
\includegraphics[height=3.0in, trim={0.5cm 0cm 5cm 0.5cm}]{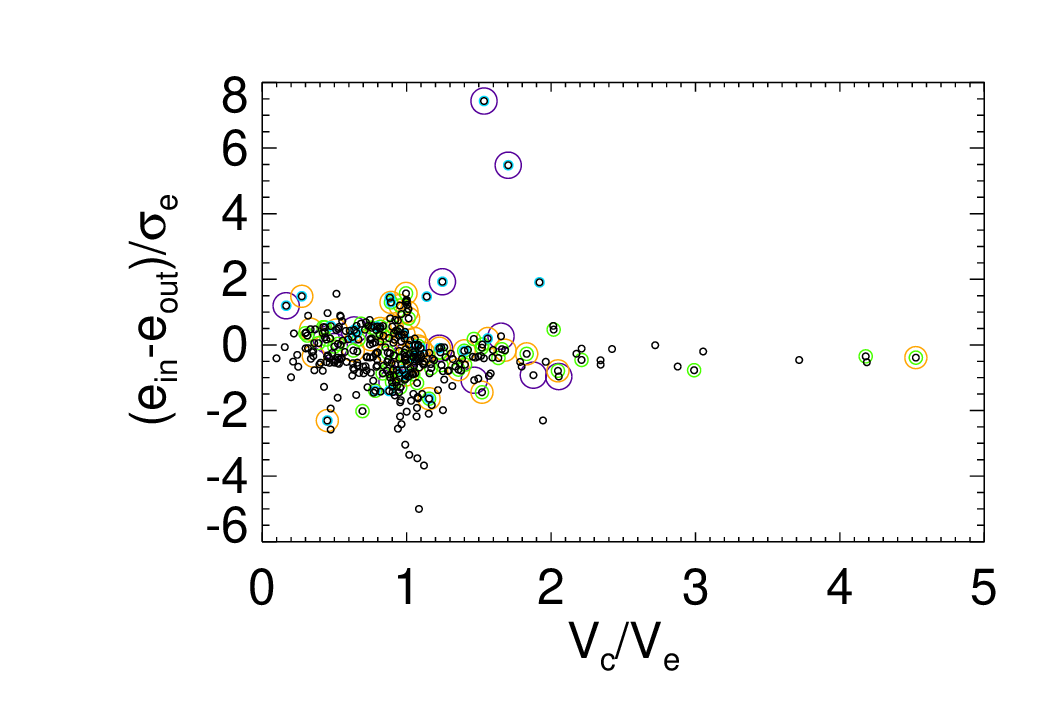}\label{fig:ediff}
\caption{A comparison of the input and output eccentricities for the 330 simulated, transit-only fits using our new eccentricity parameterization, described in \S \ref{sec:esim}, represented in black. We highlight several classes of systems one might expect to be problematic: those that are marginally mixed ($50 < T_Z < 200$ and $1.3 < R_Z < 1.5$) with a purple ring, single transit fits with a blue ring, grazing geometry ($1-R_P/R_* < b < 1+R_P/R_*$) with a green ring, and in the bottom 10\% of SNR with an orange ring. 
(a) -- The output eccentricity vs the input eccentricity. The 7 outliers (> 3-$\sigma$) are highlighted with red error bars, but are not particularly concerning. Two outliers are poorly mixed single transit systems, and the five others happen to occupy an unlikely region of $e-\omega_*$ parameter space where the posteriors accurately reflect the constraint. See text for details. (b) -- The difference between the input eccentricity and the output eccentricity, normalized by the output uncertainty, plotted as a function of the velocity during transit.}
\label{fig:esim}
\end{figure*}

\begin{figure*}
\centering
\includegraphics[height=3.0in, trim={6.0cm 0cm 10.5cm 0.5cm}]{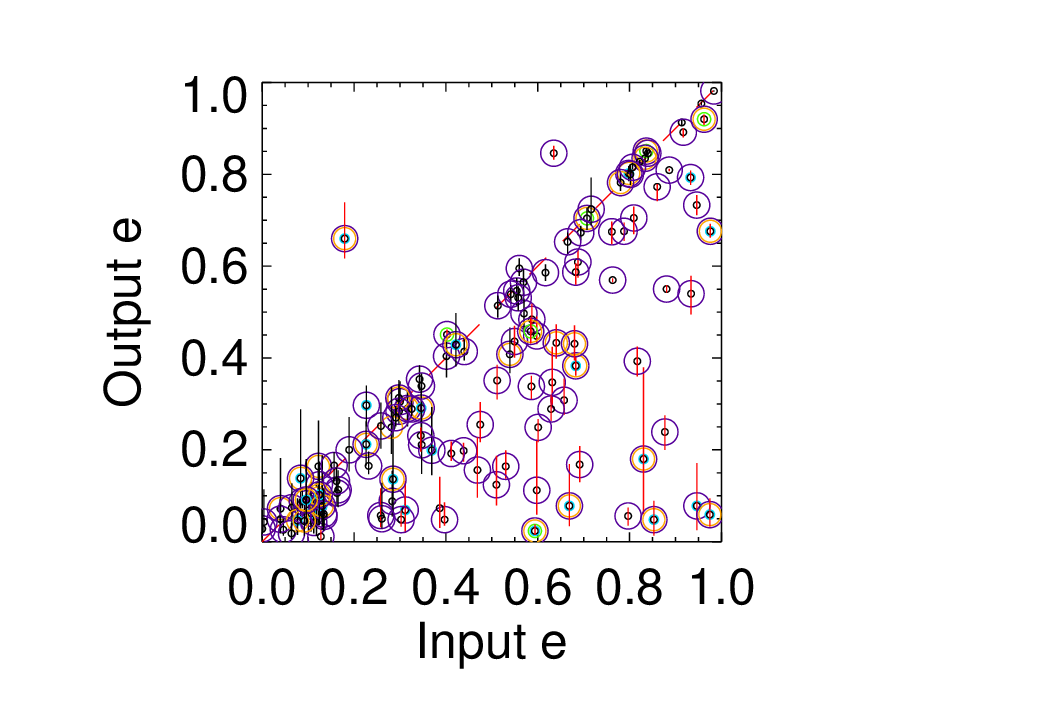}\label{fig:einveout_oldpar}
\includegraphics[height=3.0in, trim={0.5cm 0cm 5cm 0.5cm}]{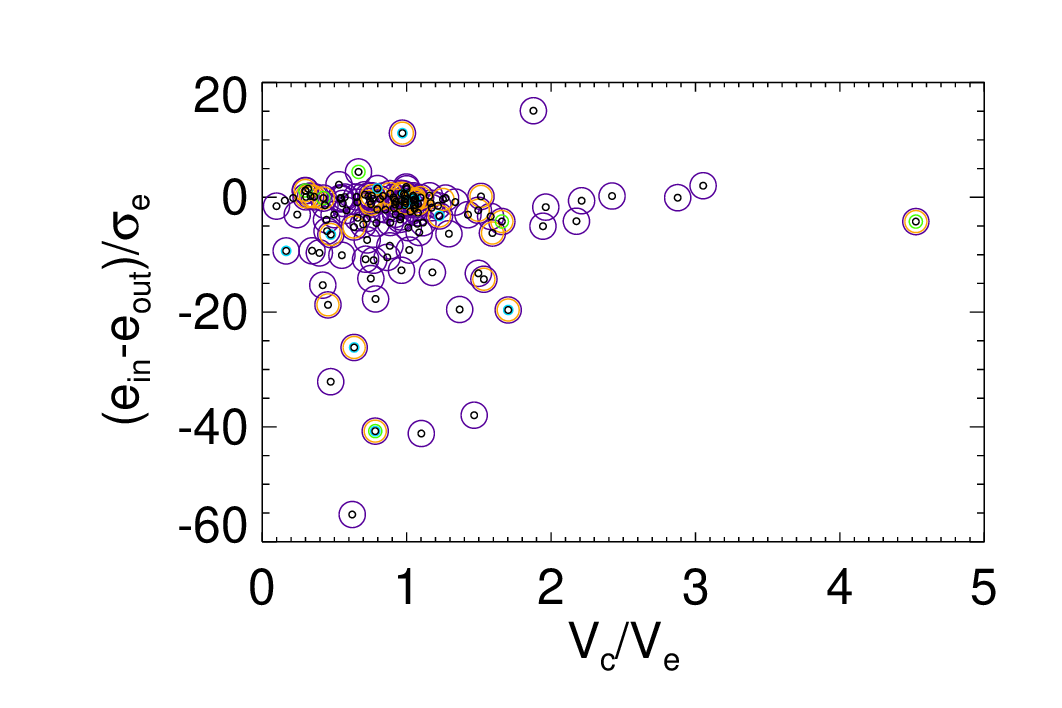}\label{fig:ediff_oldpar}
\caption{Same as Figure \ref{fig:esim}, but with the original parameterization. We see that, despite modeling the same input data with the same starting values for the same amount of time, the fits are generally far less well mixed (purple rings), far more biased, and had far more failures (missing points).}
\label{fig:esim_oldpar}
\end{figure*}

The results for eccentricity are summarized in Figures \ref{fig:esim} and \ref{fig:esim_oldpar}. We highlight systems we might expect to be problematic: where the fits were poorly mixed ($50 < T_Z < 200$ or $1.3 < R_Z < 1.5$), the signal to noise was in the bottom 10\%, the transit duration was consistent with circular (i.e., $e$ and $\omega_*$ were as degenerate as possible), a grazing geometry (i.e., inclination was poorly constrained), or a single transit (i.e., the period was poorly constrained). Typically, however, these problematic fits accurately reported an imprecise constraint.

\section{Discussion}
This new parameterization opens the door to efficient and accurate ensemble analyses of transiting systems that include eccentricity. Even in cases where we have a poor constraint on the eccentricity, we are far better off incorporating that lack of knowledge into the covariant parameters rather than fixing the eccentricity to zero and biasing the inferred planetary parameters with a non-physical model. By including the eccentricity, the global model can self-consistently link the planetary and stellar models, enabling us to use the star to constrain the planet as we outline here, but also use the transit to constrain the star \citep{Eastman:2023} without any changes to the underlying model. Having the stellar radius sets a physical scale to the system, which also allows us to model the system in the proper reference frame -- that of the target's barycenter.

\acknowledgements
Work by J.D.E. was funded by NASA ADAP 80NSSC19K1014.

The computations in this paper were run on the FASRC cluster supported by the FAS Division of Science Research Computing Group at Harvard University.


\begin{thebibliography}{}
\expandafter\ifx\csname natexlab\endcsname\relax\def\natexlab#1{#1}\fi

\bibitem[{{Anderson} {et~al.}(2011){Anderson}, {Collier Cameron}, {Hellier},
  {Lendl}, {Maxted}, {Pollacco}, {Queloz}, {Smalley}, {Smith}, {Todd},
  {Triaud}, {West}, {Barros}, {Enoch}, {Gillon}, {Lister}, {Pepe},
  {S{\'e}gransan}, {Street}, \& {Udry}}]{Anderson:2011}
{Anderson}, D.~R., {Collier Cameron}, A., {Hellier}, C., {et~al.} 2011, \apjl,
  726, L19

\bibitem[{{Bakos} {et~al.}(2007){Bakos}, {Kov{\'a}cs}, {Torres}, {Fischer},
  {Latham}, {Noyes}, {Sasselov}, {Mazeh}, {Shporer}, {Butler}, {Stefanik},
  {Fern{\'a}ndez}, {Sozzetti}, {P{\'a}l}, {Johnson}, {Marcy}, {Winn},
  {Sip{\H{o}}cz}, {L{\'a}z{\'a}r}, {Papp}, \& {S{\'a}ri}}]{Bakos:2007}
{Bakos}, G.~{\'A}., {Kov{\'a}cs}, G., {Torres}, G., {et~al.} 2007, \apj, 670,
  826

\bibitem[{{Barnes}(2007)}]{Barnes:2007}
{Barnes}, J.~W. 2007, \pasp, 119, 986

\bibitem[{{Burke}(2008)}]{Burke:2008}
{Burke}, C.~J. 2008, \apj, 679, 1566

\bibitem[{{Carter} {et~al.}(2008){Carter}, {Yee}, {Eastman}, {Gaudi}, \&
  {Winn}}]{Carter:2008}
{Carter}, J.~A., {Yee}, J.~C., {Eastman}, J., {Gaudi}, B.~S., \& {Winn}, J.~N.
  2008, \apj, 689, 499

\bibitem[{{Chen} \& {Kipping}(2017)}]{Chen:2017}
{Chen}, J., \& {Kipping}, D. 2017, \apj, 834, 17

\bibitem[{{Claret}(2017)}]{Claret:2017}
{Claret}, A. 2017, \aap, 600, A30

\bibitem[{{Dawson} \& {Johnson}(2012)}]{Dawson:2012}
{Dawson}, R.~I., \& {Johnson}, J.~A. 2012, \apj, 756, 122

\bibitem[{{Eastman} {et~al.}(2013){Eastman}, {Gaudi}, \& {Agol}}]{Eastman:2013}
{Eastman}, J., {Gaudi}, B.~S., \& {Agol}, E. 2013, \pasp, 125, 83

\bibitem[{{Eastman} {et~al.}(2023){Eastman}, {Diamond-Lowe}, \&
  {Tayar}}]{Eastman:2023}
{Eastman}, J.~D., {Diamond-Lowe}, H., \& {Tayar}, J. 2023, \aj, 166, 132

\bibitem[{{Eastman} {et~al.}(2019){Eastman}, {Rodriguez}, {Agol}, {Stassun},
  {Beatty}, {Vanderburg}, {Gaudi}, {Collins}, \& {Luger}}]{Eastman:2019}
{Eastman}, J.~D., {Rodriguez}, J.~E., {Agol}, E., {et~al.} 2019, arXiv
  e-prints, arXiv:1907.09480

\bibitem[{{Ford} {et~al.}(2008){Ford}, {Quinn}, \& {Veras}}]{Ford:2008}
{Ford}, E.~B., {Quinn}, S.~N., \& {Veras}, D. 2008, \apj, 678, 1407

\bibitem[{{Gilbert}(2022)}]{Gilbert:2022}
{Gilbert}, G.~J. 2022, \aj, 163, 111

\bibitem[{{Kipping}(2010)}]{Kipping:2010}
{Kipping}, D.~M. 2010, \mnras, 408, 1758

\bibitem[{{Kipping} {et~al.}(2012){Kipping}, {Dunn}, {Jasinski}, \&
  {Manthri}}]{Kipping:2012}
{Kipping}, D.~M., {Dunn}, W.~R., {Jasinski}, J.~M., \& {Manthri}, V.~P. 2012,
  \mnras, 421, 1166

\bibitem[{{Lucy} \& {Sweeney}(1971)}]{Lucy:1971}
{Lucy}, L.~B., \& {Sweeney}, M.~A. 1971, \aj, 76, 544

\bibitem[{{Mayo} {et~al.}(2018){Mayo}, {Vanderburg}, {Latham}, {Bieryla},
  {Morton}, {Buchhave}, {Dressing}, {Beichman}, {Berlind}, {Calkins}, {Ciardi},
  {Crossfield}, {Esquerdo}, {Everett}, {Gonzales}, {Hirsch}, {Horch}, {Howard},
  {Howell}, {Livingston}, {Patel}, {Petigura}, {Schlieder}, {Scott}, {Schumer},
  {Sinukoff}, {Teske}, \& {Winters}}]{Mayo:2018}
{Mayo}, A.~W., {Vanderburg}, A., {Latham}, D.~W., {et~al.} 2018, \aj, 155, 136

\bibitem[{{Seager} \& {Mall{\'e}n-Ornelas}(2003)}]{Seager:2003}
{Seager}, S., \& {Mall{\'e}n-Ornelas}, G. 2003, \apj, 585, 1038

\bibitem[{{Tayar} {et~al.}(2022){Tayar}, {Claytor}, {Huber}, \& {van
  Saders}}]{Tayar:2022}
{Tayar}, J., {Claytor}, Z.~R., {Huber}, D., \& {van Saders}, J. 2022, \apj,
  927, 31

\bibitem[{{Thompson} {et~al.}(2018){Thompson}, {Coughlin}, {Hoffman},
  {Mullally}, {Christiansen}, {Burke}, {Bryson}, {Batalha}, {Haas},
  {Catanzarite}, {Rowe}, {Barentsen}, {Caldwell}, {Clarke}, {Jenkins}, {Li},
  {Latham}, {Lissauer}, {Mathur}, {Morris}, {Seader}, {Smith}, {Klaus},
  {Twicken}, {Van Cleve}, {Wohler}, {Akeson}, {Ciardi}, {Cochran}, {Henze},
  {Howell}, {Huber}, {Pr{\v s}a}, {Ram{\'{\i}}rez}, {Morton}, {Barclay},
  {Campbell}, {Chaplin}, {Charbonneau}, {Christensen-Dalsgaard}, {Dotson},
  {Doyle}, {Dunham}, {Dupree}, {Ford}, {Geary}, {Girouard}, {Isaacson},
  {Kjeldsen}, {Quintana}, {Ragozzine}, {Shabram}, {Shporer}, {Silva Aguirre},
  {Steffen}, {Still}, {Tenenbaum}, {Welsh}, {Wolfgang}, {Zamudio}, {Koch}, \&
  {Borucki}}]{Thompson:2018}
{Thompson}, S.~E., {Coughlin}, J.~L., {Hoffman}, K., {et~al.} 2018, \apjs, 235,
  38

\bibitem[{{Tingley} \& {Sackett}(2005)}]{Tingley:2005}
{Tingley}, B., \& {Sackett}, P.~D. 2005, \apj, 627, 1011

\bibitem[{{Van Eylen} \& {Albrecht}(2015)}]{VanEylen:2015}
{Van Eylen}, V., \& {Albrecht}, S. 2015, \apj, 808, 126

\bibitem[{{Winn}(2010)}]{Winn:2010}
{Winn}, J.~N. 2010, arXiv e-prints, arXiv:1001.2010

\end{thebibliography}

\appendix
\section{Using $L$ to choose the solution for $\MakeLowercase{e}$}
\label{sec:elparam}

As discussed in \S \ref{sec:edegeneracy}, we parameterize $e$ and $\omega_*$ as $V_c/V_e$, \lsinw \ and \lcosw. Because $e$ has two solutions for any given pair of $V_c/V_e$ and $\omega_*$, we must choose between them in a way that does not a priori favor one over the other and bias the priors away from uniform. We concluded that adding an extra parameter, $S$, resulted in faster convergence times, but here we detail investigations we did to using the degenerate parameter $L$ to avoid adding an unnecessary parameter.

When $\left(\lsinw\right)^2 + \left(\lcosw\right)^2 = L^2 > 0.5$, we choose the solution with a negative sign. Otherwise, we choose the solution with the positive sign. This boundary is chosen to ensure that both solutions for $e$ are a priori equally likely. Unfortunately, the covariance is not as simple as Figure \ref{fig:vcve} would suggest. The covariance between $V_c/V_e$ is, by construction, well behaved with respect to both \lsinw \ and \lcosw, but the covariance between \lsinw \ and \lcosw \ is more complex. 

\begin{figure*}
\centering
\includegraphics[height=2.0in, trim={5cm 0cm 7cm 0.5cm}]{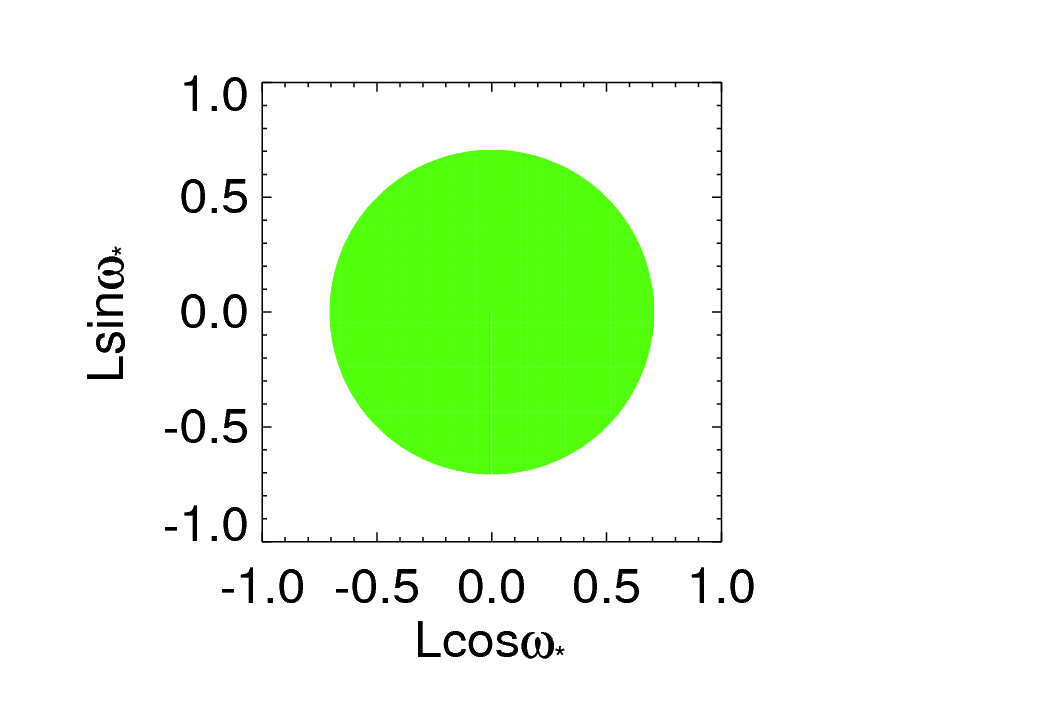}\label{fig:vcve0.5}
\includegraphics[height=2.0in, trim={5cm 0cm 7cm 0.5cm}]{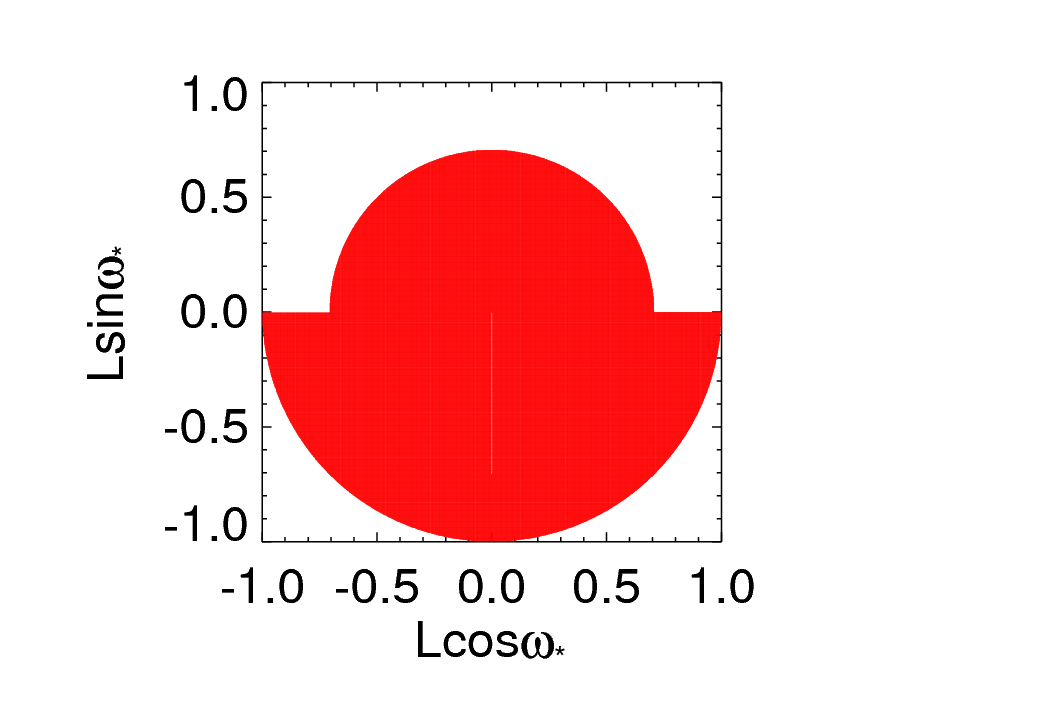}\label{fig:vcve1.0}
\includegraphics[height=2.0in, trim={5cm 0cm 7cm 0.5cm}]{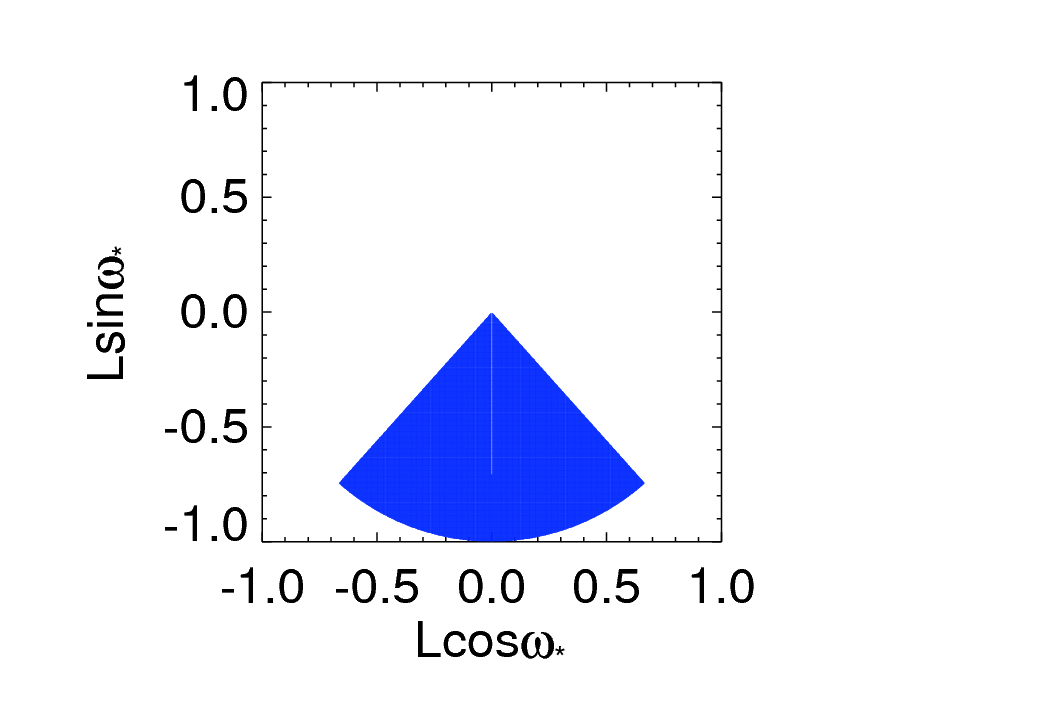}\label{fig:vcve1.5}
\caption{The covariance between $L\sin{(\omega_{*})}$ and $L\cos{(\omega_{*})}$, for a $V_c/V_e = 0.5$ (panel (a) in green), $V_c/V_e = 1.0$ (panel (b) in red), and $V_c/V_e = 1.5$ (panel c in blue). The whitespace in these plots correspond to non-physical models that are rejected. We take the negative solution to the quadratic when $\left(\lsinw\right)^2 + \left(\lcosw\right)^2 = L^2 > 0.5$ and the positive solution otherwise.}
\label{fig:lcoswlsinw}
\end{figure*}

\begin{figure*}
\centering
\includegraphics[height=2.0in, trim={5cm 0cm 7cm 0.5cm}]{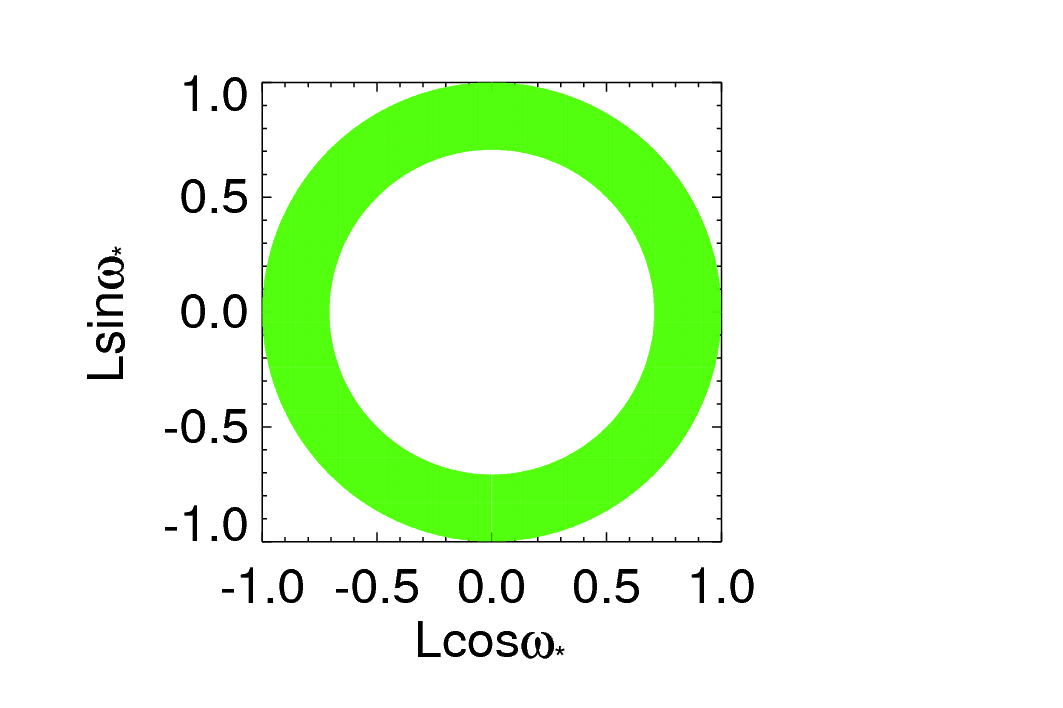}\label{fig:vcve0.5bad}
\includegraphics[height=2.0in, trim={5cm 0cm 7cm 0.5cm}]{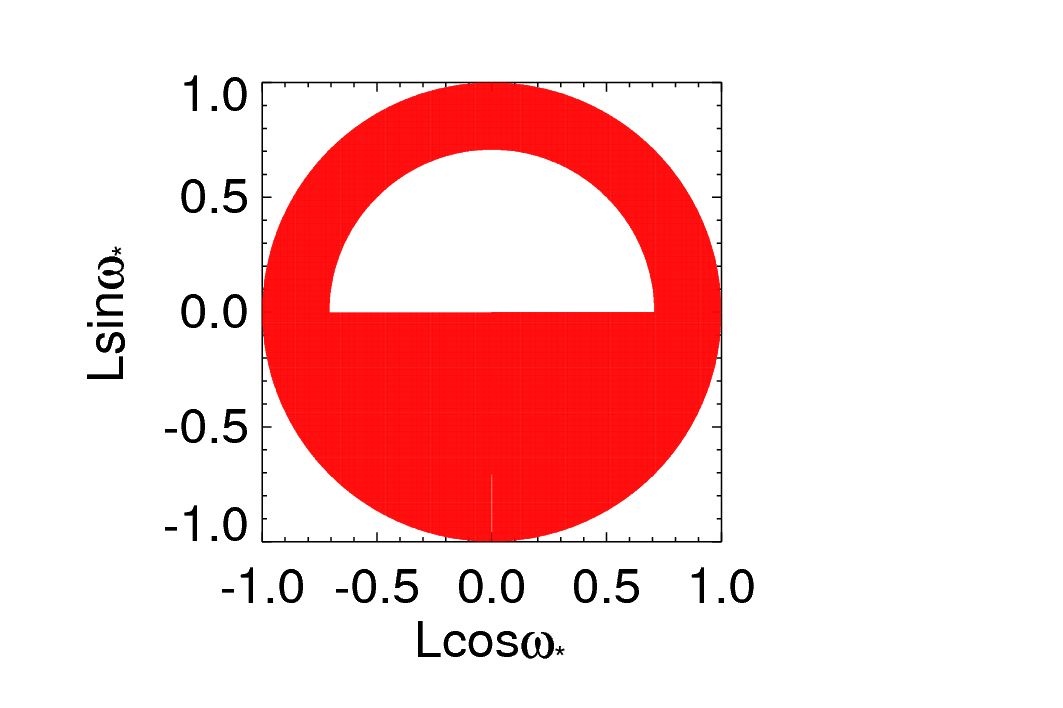}\label{fig:vcve1.0bad}
\includegraphics[height=2.0in, trim={5cm 0cm 7cm 0.5cm}]{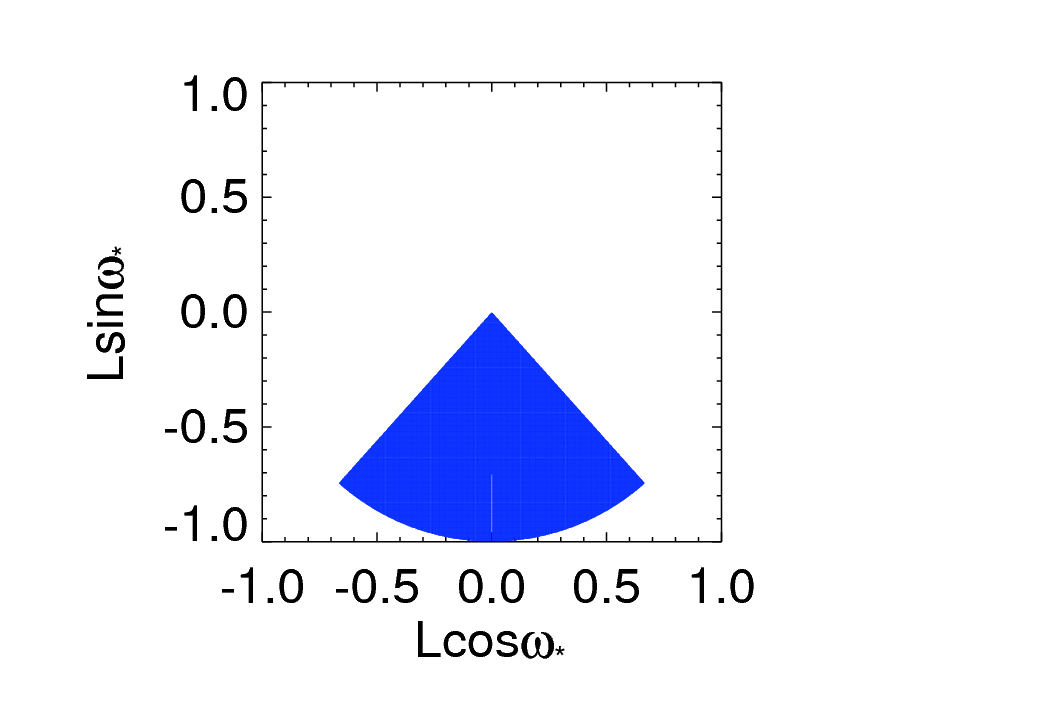}\label{fig:vcve1.5bad}
\caption{Same as Figure \ref{fig:lcoswlsinw}, but with the opposite definition of $L$ -- we take the positive solution to the quadratic when $\left(\lsinw\right)^2 + \left(\lcosw\right)^2 = L^2 > 0.5$ and the negative solution otherwise. Note these hollow regions are diabolical for DE-MCMC to sample, and so this definition is strongly discouraged.}
\label{fig:lcoswlsinwbad}
\end{figure*}

Figures \ref{fig:lcoswlsinw} and \ref{fig:lcoswlsinwbad} show this covariance for our three representative values of $V_c/V_e$. The shaded regions have the same meanings as before -- green corresponds to $V_c/V_e=0.5$, red corresponds to $V_c/V_e=1.0$, and blue corresponds to $V_c/V_e=1.5$ -- but since each panel is at a constant $V_c/V_e$, there are no contours to show. The shaded regions are degenerate with the duration constraint from the transit, while the white-space corresponds to non-physical solutions ($e$ is either imaginary, negative, or greater than 1).

The difference between the two figures is that, when $L^2 < 0.5$, figure \ref{fig:lcoswlsinw} uses the negative solution to solve for $e$ and figure \ref{fig:lcoswlsinwbad} uses the positive solution. By comparing the two figures, we see that this seemingly arbitrary choice is actually critically important. By using the negative solution of the quadratic, the physical solutions are contiguous and relatively well behaved. If we were to use the positive solution instead, the physical solutions when $V_c/V_e \leq 1$ would form a ring in \lsinw--\lcosw \ space. As the DE-MCMC algorithm samples that ring, it would tend to propose many steps in the non-physical region inside the ring -- similar to the problem with the \secosw, \sesinw \ parameterization we were trying to avoid in the first place.f

However, even with the better choice of $L$, a single test-case for Kepler-75b converged about two times faster when we introduce an additional sign parameter, $S$, instead. We bound $0 \leq S < 2$, and when {\tt floor}$(S) <1$, we choose the positive solution. Otherwise, we choose the negative solution. This is likely to be generally better due to the sharp edges in \lcosw--\lsinw \ space when $L$ is fit, even though adding an additional parameter may seem wasteful. Still, this tradeoff is relatively unexplored, and there may be cases where using $L$ to choose the sign of the solution is better or slight tweaks along these lines could further improve performance.

\end{document}